\documentstyle[12pt]{article}

\newcommand{\be}{\begin{equation}}
\newcommand{\ee}{\end{equation}}
\newcommand{\bea}{\begin{eqnarray}}
\newcommand{\eea}{\end{eqnarray}}

\newcommand{\vekt}[1]{{\bf #1}}
\newcommand{\order}{{\cal O}}
\newcommand{\Ang}{{\rm \AA}}
\newcommand{\Kelvin}{^\circ{\rm K}}
\newcommand{\kB}{k_{\rm B}}
\newcommand{\half}{{1 \over 2}}
\newcommand{\gsim}{\,\raisebox{-1.0ex}{$\stackrel{\displaystyle
>}{\sim}$}\,}
\newcommand{\lsim}{\,\raisebox{-1.0ex}{$\stackrel{\displaystyle
<}{\sim}$}\,}
\newcommand{\grad}{\nabla}
\newcommand{\const}{{\rm const.}}
\newcommand{\etal}{{\em et al.\ }}

\newcommand{\Tv}{T_{\rm v}}

\newcommand{\Tc}{T_{\rm c}}
\newcommand{\Rc}{R_{\rm c}}
\newcommand{\Tco}{T_{\rm c0}}
\newcommand{\TCG}{T^{\rm \scriptscriptstyle CG}}
\newcommand{\TCGc}{T^{\rm \scriptscriptstyle CG}_{\rm c}}
\newcommand{\TCGv}{T^{\rm \scriptscriptstyle CG}_{\rm v}}
\newcommand{\TCGst}{T^{\rm \scriptscriptstyle CG}_*}
\newcommand{\zCG}{z^{\rm \scriptscriptstyle CG}}
\newcommand{\zCGc}{z^{\rm \scriptscriptstyle CG}_{\rm c}}
\newcommand{\zCGv}{z^{\rm \scriptscriptstyle CG}_{\rm v}}
\newcommand{\zCGst}{z^{\rm \scriptscriptstyle CG}_*}
\newcommand{\muCG}{\mu^{\rm \scriptscriptstyle CG}}
\newcommand{\muXY}{\mu_{\rm \scriptscriptstyle XY}}
\newcommand{\muGL}{\mu_{\rm \scriptscriptstyle GL}}

\newcommand{\Lams}{\Lambda_{\rm s}}
\newcommand{\lamL}{\lambda_{\rm L}}

\newcommand{\He}{{\rm He}}
\newcommand{\mpar}{m_\|}
\newcommand{\me}{m_{\rm e}}
\newcommand{\ntwoD}{n_{\rm 2D}}

\newcommand{\ta}{\tilde{a}}
\newcommand{\tS}{\tilde{S}}
\newcommand{\tH}{\tilde{H}}
\newcommand{\hr}{\hat{r}}
\newcommand{\hR}{\hat{R}}
\newcommand{\hF}{\hat{F}}
\newcommand{\vr}{\vekt{r}}
\newcommand{\vn}{\vekt{n}}
\newcommand{\vo}{\vekt{0}}
\newcommand{\vk}{\vekt{k}}
\newcommand{\vR}{\vekt{R}}
\newcommand{\vA}{\vekt{A}}
\newcommand{\vF}{\vekt{F}}

\title{Possible first order transition in the two-dimensional Ginzburg-
Landau model induced by thermally fluctuating vortex cores}

\author{Dierk Bormann\thanks{Universit\"at Augsburg, Institut f\"ur Physik,
Memminger Str.\ 6, D-86135 Augsburg} \ and
        Hans Beck\thanks{Universit\'e de Neuch\^atel, Institut de Physique,
Rue A.-L. Breguet 1, CH-2000 Neuch\^atel}
       }

\date{}

\begin{document}

\maketitle

\begin{abstract}
We study the two-dimensional Ginzburg-Landau model of a neutral superfluid
in the vicinity of the vortex unbinding transition.
The model is mapped onto an effective interacting vortex gas by a systematic
perturbative elimination of all fluctuating degrees of freedom (amplitude
{\em and} phase of the order parameter field) except the vortex positions.
In the Coulomb gas descriptions derived previously in the literature,
thermal amplitude fluctuations were neglected altogether.
We argue that, if one includes the latter, the vortices still form a two-
dimensional Coulomb gas, but the vortex fugacity can be substantially
raised.
Under the assumption that Minnhagen's generic phase diagram of the two-
dimensional Coulomb gas is correct, our results then point to a first order
transition rather than a Kosterlitz-Thouless transition, provided the
Ginzburg-Landau correlation length is large enough in units of a microscopic
cutoff length for fluctuations.
The experimental relevance of these results is briefly discussed.

\vspace{.5cm}\noindent
\end{abstract}

\newpage

\section{Introduction}

The critical behavior of two-dimensional (2D) systems with a continuous
internal symmetry has been a most puzzling problem for a long time.
Simple physical realizations are superfluid or superconducting thin films,
which on a phenomenological level can both be described by a complex order
parameter field $\Psi$, the boson or Cooper pair ``condensate wavefunction".
The internal symmetry is then a U(1) gauge symmetry acting on the phase of
$\Psi$.

Early theoretical work showed that the ``usual" criterion for superfluid
order in 3D bulk systems \cite{ODLRO}, namely long-range order (LRO) of the
field $\Psi$, is not satisfied in 2D films at any finite temperature
\cite{2DLRO}.
The reason are low-energy phase fluctuations, the Goldstone modes related to
the broken gauge symmetry, leading to $\Psi$ correlations which decay
algebraically to zero with distance.
However, it was quickly realized \cite{Las68} that this ``quasi"-LRO was
sufficient to insure superfluidity.
The correct criterion for superfluidity turned out to be rather a
nonvanishing stiffness with respect to long-wavelength phase fluctuations
(the {\em helicity modulus} $\Upsilon$ \cite{FBJ73}) than true LRO in
$\Psi$.

As a matter of fact, experimentally even very thin $^4$He films of a {\em
fraction} of an atomic layer showed clear signatures of superfluidity (see
refs.\ in \cite{BR80}).
2D superconductivity was predicted \cite{BMO79,HN79} for sufficiently
``dirty" samples (high normal sheet resistance) and observed in continuous
and granular thin films (see e.g.\ \cite{HF80,SCfilms,LFM90,VLV91}).

The nature of the transition from this superfluid phase to a high-
temperature phase with exponentially decaying $\Psi$ correlations was
clarified by Berezinski\u{\i} \cite{Ber71} and Kosterlitz and Thouless (KT)
\cite{KT72,KT73}.
They realized the essential role of vortices, i.e.\ phase singularities with
nonvanishing winding number (``vorticity").
The vortices interact logarithmically at large distances, thus at low
temperatures they appear as bound pairs of zero total vorticity only which
do not change the algebraic decay of the $\Psi$ correlations.
However, at some finite temperature $\Tv$, the largest vortex pairs in the
system start to dissociate (unbind) by a collective screening mechanism, and
the free vortices lead to an exponential decay of the correlations.
(We use the notation $\Tv$ here in order to avoid confusion with a
``conventional" critical point which will appear later.)

Usual phenomenological models of superconducting or superfluid films are
the Ginzburg-Landau (GL) model or the XY (``planar rotator") model
\cite{HN79,Nel83}, respectively (see also section 2).
The GL model is generally believed to provide an appropriate description of
superconductors close to the bulk transition since there is a microscopic
derivation in the framework of Gor'kov theory (see e.g.\ refs.\
\cite{BCSGL}).
The XY model may be considered as the ``phase only" limit of the GL model,
in which the modulus of $\Psi$ is fixed and only its phase is allowed to
fluctuate \cite{Hal79}.
Together with the idea of thermally excited vortex loops, it has been
succesfully employed to describe the superfluid transition of $^4\He$ in 3D
\cite{vortexloops} and 2D \cite{HN79,Nel83}, but there is no comparable
microscopic justification as for the GL model in the case of
superconductors.

Since the KT transition is essentially caused by the interacting vortex
system only, it appears in its purest form in a corresponding model system
of pointlike particles in 2D with a logarithmic interaction, the 2D Coulomb
gas (2DCG) model (for a review see e.g.\ \cite{Min87}).
This model depends on two independent, dimensionless parameters, a
temperature $\TCG$ and a fugacity $\zCG$ of the particles.
The GL and XY models may then be thought of as particular realizations of
the CG model, represented by particular $\zCG(\TCG)$ lines \cite{Min87}.

Based on the ideas of Kosterlitz and Thouless, a systematic renormalization
group (RG) theory of the KT transition has been developed, usually starting
from an XY or CG type description \cite{KTRG,JKK77}.
In the CG picture, the KT RG equations are flow equations in the $\zCG$-
$\TCG$ plane; the $\zCG(\TCG)$ line of a particular realization serves as
initial condition of the RG flow.
In the RG framework it could for instance be shown \cite{NK77} that the
helicity modulus has a finite value $\Upsilon(T \to \Tv^{(-)}) \neq 0$ just
below the transition and then drops {\em discontinuously} to zero.
Furthermore, the ratio $2\pi \Upsilon(T)/\kB T$ tends to the {\em universal}
value 4 at the transition ($T \to \Tv^{(-)}$).
This result may also be expressed in terms of pure CG quantities: general
hydrodynamic arguments \cite{MW81} show that the quantity $(\epsilon_0(\TCG)
\TCG)^{-1}$ ($\epsilon_0$ being the $k \to 0$ limit of the CG dielectric
function) corresponds {\em exactly} to the above-mentioned ratio $2\pi
\Upsilon(T)/\kB T$ at all temperatures, in particular its discontinuity is
the same.

This famous ``universal jump" prediction is one of the central results of
the KT RG theory and has been verified to an impressive degree by
measurements on $^4\He$ films \cite{BR80} as well as by numerical work on XY
type models (see e.g.\ \cite{JN91} and refs.\ therein, refs.\ in
\cite{JMN93}) and CG type models \cite{CGnumer}.
Quite convincing evidence for KT universal behavior has also been obtained
for artificial superconductor networks, more specifically weakly coupled
Josephson junction arrays (JJA's) \cite{JJAs} and wire networks
\cite{JFG89}, which are both well described by XY type models (see section
2).

The situation for continuous superconducting films appears to be less clear.
Experiments on high-$\Tc$ films (YBCO \cite{LFM90,VLV91}) and dirty
conventional films (Al \cite{HF80}, In/InO \cite{SCfilms}) films provide
rather good evidence for a discontinuous jump of the helicity modulus
$\Upsilon$ at the transition, however in both cases they seem to point to
slightly {\em larger} values of this jump.
(We note that in one case --- ref.\ \cite{LFM90} --- the authors explain the
observed behavior of $\Upsilon$ in a completely different way, namely by
percolation effects due to the granularity of the film.)

Unfortunately, in the literature apparently no numerical results are
available on the critical behavior of the 2D GL model.
Usually, the 2D GL model is assumed \cite{Hal79,Nel83} to have the same
universal properties as the XY and CG models, since small amplitude
fluctuations of the order parameter $\Psi$ have been shown \cite{Pel78} to
be {\em irrelevant} in the RG sense.
These fluctuations are then neglected altogether, and one is left with a gas
of GL vortices (termed ``Ginzburg-Landau Coulomb gas" (GLCG) in the
literature \cite{MN85,Min87}; we will call it the ``bare" GLCG since
fluctuations are not yet included) whose interaction is logarithmic at large
distances and cut off roughly at the GL correlation length $\xi$.
However, it is easily seen (see subsection 2.1) that {\em right at} the
expected vortex unbinding transition, the amplitude fluctuations cannot be
considered as weak anymore and thus may well change the critical behavior.

A further, conceptual problem of the ``bare" GLCG is that the {\em phase
space division} $\Delta$ of a vortex is not well defined without additional
arguments \cite{Min87}.
To get a rough idea of the size of $\Delta$, one may argue that neglecting
fluctuations at length scales $\lsim \xi$ corresponds to a lattice
regularization with lattice spacing $\sim \xi$.
Consequently, the vortex phase space division is expected to be $\sim
\xi^2$, since a vortex can be located at any of the plaquettes of the
lattice.
A more systematic approach to this question will be discussed at the end of
this paper, in subsection 4.4.

The above qualitative argument implicitly assumes that the only length scale
which enters the problem is $\xi$, in particular the $\zCG(\TCG)$ line
should be {\em independent} of any microscopic cutoff length for
fluctuations.
In contrast to this, in the present paper we study precisely the influence
of small-wavelength ($\lsim \xi$) fluctuations on the transition.
In doing so, we still assume that a description of the transition in terms
of an effective interacting vortex gas is justified.
We then take  all thermal fluctuations systematically into account by a
perturbative elimination of all degrees of freedom except the vortex
positions.
Our results will indicate that these fluctuations strongly {\em increase}
the density of vortices at the transition or, in other words, they shift up
the $\zCG(\TCG)$ line.

Or, already KT in their original paper \cite{KT73} noted that their
approximations are justified only when the vortex system is dilute enough at
the transition, i.e.\ when $\zCGv = \zCG(\TCGv)$ is small enough.
Outside the region of small $\zCGv$, universality will hold only as far as
the RG flow remains KT like in a topological sense.
On the other hand, Minnhagen \cite{Min85} (see also \cite{Min87,TK88,MW89})
investigated the {\em whole} phase diagram of the 2D CG model by an
extension to larger $\zCG$ of the KT-Young self-consistent screening
procedure \cite{KT73,You78} of deriving the RG equations.
Minnhagen's modified RG equations are nonholonomous integro-differential
equations which in the small $\zCG$ limit reduce to the KT equations.
Moreover, for small enough $\zCG$ his corrections to the latter are {\em
irrelevant} in the RG sense, i.e.\ Minnhagen's equations reproduce the same
universal properties in this region.
At larger values of $\zCG$ however, his equations predict a first order
transition line which ends at some critical point $(\zCGc,\TCGc) \approx
(0.029,0.204)$.
The KT line joins the first order line smoothly from below at some point
$(\zCGst,\TCGst) \approx (0.054,0.144)$ (fig.\ \ref{CG1}, qualitatively
adapted from \cite{MW89}).
The superfluid transition is KT like up to this point and first order
further above.

The first order part probably may still be interpreted as a vortex unbinding
transition, which in contrast to the KT transition involves the simultaneous
dissociation of a {\em finite fraction} of all bound vortex pairs in the
system.
Its characteristics are nonuniversal, for instance the jump of the above-
mentioned quantity $(\epsilon_0(\TCG) \TCG)^{-1}$ depends on $\TCGv$ in the
first order part of the transition, and it is {\em larger} than the KT value
4.

In fig.\ \ref{CG1}, we have included the $\zCG(\TCG)$ lines representing the
XY model and the ``bare" GLCG, where for the latter we used the functional
form proposed in \cite{WM88} (see also subsection 2.3).
The XY line definitely lies in the KT regime, there is thus no contradiction
between the Minnhagen theory and the numerically well established fact (see
references in \cite{JN91} and \cite{JMN93}) that the XY model displays the
universal properties of a KT transition.
The GLCG line intersects both the KT and the first order lines, but the
superfluid transition is still KT like.
Our aim here is to argue that inclusion of fluctuations may shift the
$\zCG(\TCG)$ line even further up in the first order regime.

However, the Minnhagen scenario must still be considered as somewhat
speculative.
Some effort has been made in the past (see \cite{JMN93,Mil93} and references
therein) to decide whether a first order transition exists in a modified XY
model with a ``truncated" cosine interaction between neighboring phase
angles (which presumably also corresponds to a {\em higher} $\zCG(\TCG)$
line than the XY one in fig.\ \ref{CG1}) but the conclusions obtained by
different authors are contradictory.
Perhaps the strongest case in favour of the Minnhagen scenario could be made
by Jonsson \etal \cite{JMN93}.
By a finite-size scaling analysis, they established the existence of a
critical point {\em above} the KT line, which they identify with
$(\zCGc,\TCGc)$.

In summary, in this paper we try to treat the mapping of the 2D GL model on
the corresponding interacting vortex gas in a somewhat more complete and
systematic fashion than it has been done so far in the literature.
Thereby we find that fluctuations strongly enhance the vortex fugacity at
the transition, which may drive the transition first order, provided the
Minnhagen szenario is correct.

The paper is organized as follows: in section 2, we start with some general
remarks about the GL description of superconducting films and networks, and
also define our modified notion of the ``Ginzburg-Landau vortex gas"
(GLVG).
The details of the main calculation are contained in sections 3 and 4.
The idea is to first investigate some kind of ``saddle point" configuration
of the field for given vortex positions (section 3), which corresponds
roughly to the ``bare" GLCG discussed previously in the literature.
In a second step, we include thermal fluctuations around this configuration
in a gaussian approximation (section 4) and derive, as our central result,
the $\zCG(\TCG)$ relation for the GLVG.
Section 5 contains a short summary and conclusions.

\section{The effective Ginzburg-Landau vortex gas}

\subsection{Ginzburg-Landau description of superconducting films}

The starting point of this paper is the GL free energy functional of a
complex order parameter field $\Psi$ in 2D,
\be
  {\cal H}[\Psi] = \int d^2 {\vr} \left\{ \alpha |\Psi|^2
  + {\beta \over 2}|\Psi|^4 + \gamma|\grad \Psi|^2 \right\} \ .
\label{ZA}
\ee
A {\em physical} system can be considered as 2D, if the thickness of the
film is smaller than some minimum wave length of $\Psi$ fluctuations, such
that variations of $\Psi$ perpendicularly to the plane are negligible.
Note that we also omitted any coupling to the magnetic vector potential
$\vA$, i.e.\ (\ref{ZA}) actually describes a {\em neutral} superfluid.
Nevertheless, as argued in the introduction such a local free energy is
probably more appropriate for superconducting films than for $^4\He$ films.
At the end of this section, we will indicate under which circumstances the
omission of a coupling to $\vA$ can be justified.

$\cal H$ is used to define statistical-mechanical quantities like the
partition sum
\be
  Z \propto \int {\cal D}\Psi e^{-{\cal H}[\Psi]/\kB T} \ .
\label{ZB}
\ee
The coefficients $\alpha, \beta, \gamma$ in $\cal H$ are in general
temperature dependent quantities: as usual, we assume that $\alpha$ vanishes
at some ``mean field" critical temperature $\Tco$, and that $\alpha < 0$ for
$T < \Tco$.
Furthermore, for (\ref{ZB}) to be defined, necessarily $\beta, \gamma > 0$.
There is some freedom in the interpretation of the parameters $\alpha,
\beta, \gamma$ and the order parameter field $\Psi$; usually, one
chooses $\gamma = \hbar^2/2\mpar$ where $\mpar$ is the effective in-plane
mass of the carriers (electrons or holes) and interpretes $|\Psi|^2$ as the
local density of carriers in the condensate.
If we disregard boundaries, (\ref{ZA}) assumes its minimum for the
homogeneous field configuration $\Psi_\infty \equiv \sqrt{|\alpha|/\beta}$
and for any other configuration which differs from $\Psi_\infty$ by an
arbitrary, spatially constant phase factor.

For later convenience and clarity we introduce a dimensionless order
parameter field $\psi := \sqrt{\beta/|\alpha|} \Psi$ normalized such that
$\psi_\infty \equiv 1$.
With
\be
  K := \frac{2 \gamma |\alpha|}{\beta \kB T} \ , \qquad
  \xi := \sqrt{\frac{\gamma}{|\alpha|}}
\label{ZC}
\ee
and
\be
  H[\psi] = \int d^2 {\vr} \left\{ \frac{1}{2\xi^2} \left( 1-|\psi|^2
            \right)^2 + |\grad \psi|^2 \right\}
\label{ZD}
\ee
we can then rewrite the exponent of (\ref{ZB}) as
\be
\frac{{\cal H}[\Psi]}{\kB T} = \frac{K}{2} H[\psi] + \const \ .
\label{ZE}
\ee
``const." stands for a $\psi$-independent term which is irrelevant for the
thermodynamics.
$\xi$ is the GL ``correlation" length which measures the length scale on
which $\psi$ relaxes to $\psi_\infty = 1$ away from a perturbation.
We will see later that it is also the correlation length for thermal
fluctuations of the amplitude $|\psi|$.
$K$ measures the stiffness of the $\psi$ field over temperature and its
inverse will later play the role of an effective statistical-mechanical
temperature.
Note that $1/K$ and $\xi$ both {\em diverge} as $T$ approaches $\Tco$ from
below, since $|\alpha| \to 0$.
Both effects tend to enhance fluctuations, so if the KT transition is not
preceded by some other transition to a disordered high-temperature phase,
there must be a vortex unbinding transition at some temperature $\Tv $ {\em
below} $ \Tco$.

Expression (\ref{ZD}) looks as if $\xi$ were the only length scale in the
problem and could be eliminated by a simple rescaling of all lengths, i.e.\
of $\vr$.
This is in fact the case in usual GL theory \cite{BCSGL}.
However, in order to obtain well-defined, finite results for statistical-
mechanical quantities like the partition sum (\ref{ZB}) or $\psi$
correlation functions, we have to limit the number of degrees of freedom by
a UV regularization which introduces a second length scale.
For definiteness, we may for instance put the model (\ref{ZD}) on a square
lattice with spacing $a$,
\be
  H[\psi] = \frac{a^2}{2\xi^2} \sum_i \left( 1-|\psi_i|^2 \right)^2
            + \sum_{\langle ij \rangle} |\psi_i - \psi_j|^2 \ ,
\label{ZF}
\ee
where $\sum_{\langle ij \rangle}$ denotes a sum over all pairs of nearest
neighbor sites.
Alternatively, we may supplement the continuum version (\ref{ZD}) by the
corresponding momentum cutoff prescription (i.e., all momenta in the Fourier
representation of (\ref{ZD}) restricted to the first Brillouin zone $[-
\frac{\pi}{a},\frac{\pi}{a}]^2$).
We will assume that both prescriptions are essentially equivalent and use
them in parallel.

Expression (\ref{ZF}) clearly shows that $\xi/a$ is a second (besides $K$)
dimensionless parameter which enters the model.
In the limit $\xi \ll a$, the first term in (\ref{ZF}) suppresses amplitude
fluctuations of the order parameter away from the value $|\psi_i| \equiv 1$
and (\ref{ZF}) reduces to the XY hamiltonian,
\be
  H_{\rm XY}[\psi] = \sum_{\langle ij \rangle} |e^{i\theta_i} -
e^{i\theta_j}|^2
        = 2 \sum_{\langle ij \rangle} \left( 1 - \cos( \theta_i - \theta_j )
\right) \ ,
\label{ZG}
\ee
where $\theta_i$ is the phase of $\psi_i$.
In this respect, the XY model is the ``phase only" $\xi \to 0$ limit of the
GL model.
As mentioned in the introduction, its predictions agree well with
measurements on
Josephson junction arrays (JJA's) and superconducting wire networks.
 From our point of view, the reason for this success is that in these
artificial networks the spacing of the underlying lattices provide a
macroscopic cutoff length $a$ which can be tuned independently of $\xi$.
{\em Weakly} coupled networks \cite{JJAs,JFG89} which are preferred
experimentally since their transition temperature is well seperated from the
bulk transition then always lie in the regime $\xi \ll a$.
Note that in a model of type (\ref{ZF}) for weakly coupled JJA's, $\xi$ is
{\em not} equal to the bulk GL correlation length but usually much smaller
(as a matter of fact, $\xi^2/a^2$ is essentially proportional to the
Josephson coupling energy between two grains over the condensate energy of a
grain).
Consequently, there is no contradiction to the fact that usually $a$ is
smaller than the bulk $\xi$ in these systems.
\newline
The main points we want to stress now are the following:

\begin{enumerate}

\item
In continuous superconducting films (as opposed to networks), $\xi$ is {\em
not} independent of $a$.
In fact, it always obeys the inequality $\xi \gsim a$.

\item
If $\xi \gsim a$, thermal amplitude fluctuations are quite strong close to
the transition.
\end{enumerate}
In the remainder of this paper, we will then analyse in detail the effect of
these fluctuations and discuss in which way they may change the critical
behavior.

In order to demonstrate point 2., we disregard for the moment the phase
degrees of freedom and estimate the local fluctuations of the amplitude in a
gaussian approximation.
An expansion of the potential term in the hamiltonian (\ref{ZD}) shows that
amplitude fluctuations $\delta |\psi| = |\psi| - 1$ have a ``mass"
$2/\xi^2$.
Therefore their mean square value at some ``temperature" $1/K$ may be
estimated as
\bea
  \lefteqn{ \frac{ \langle (\delta|\psi|)^2
    \rangle }{ \langle |\psi| \rangle^2 }
    \approx \frac{1}{2 K} \int_{-\pi/a}^{\pi/a}
    \frac{d^2 \vk}{(2\pi)^2} \frac{1}{2/\xi^2 + \vk^2}
  } \nonumber \\
  & & \approx \frac{1}{8\pi K} \int _0^{4\pi/a^2}
    \frac{d(k^2)}{2/\xi^2 + k^2}
    = \frac{1}{8\pi K} \ln \left(1 + 2\pi \frac{\xi^2}{a^2} \right) \ ,
  \label{ZH}
\eea
where in the second line we just have approximated the quadratic Brilluoin
zone by a circular one of the same area $(2\pi/a)^2$.
At the presumed vortex unbinding transition, one expects the dimensionless
``inverse temperature" $K$ to be roughly of the order of 1, which implies
that for $\xi \gsim a$ the $|\psi|$ fluctuations are of the same order of
magnitude as its expectation value squared, $\langle |\psi| \rangle^2
\approx 1$.

To show that for superconducting films always $\xi \gsim a$ (point 1. above)
we first have to understand the meaning of cutoff $a$ in this context.
In fact, the Gor'kov derivation (\cite{Gor59}, see also the reviews
\cite{BCSGL}) of the GL functional (\ref{ZA}) from BCS theory does not
immediately yield a local form like (\ref{ZA}), but involves integral
kernels whose range plays the role of the cutoff
$a$.
Both $a$ and the GL correlation length $\xi$ can be calculated in this
framework and one obtains roughly
\be
  \frac{\xi}{a} \approx \frac{1}{ \sqrt{\chi (1 - T/\Tco)} } \ ,
\label{ZI}
\ee
where $\chi \approx (1 + \hbar/2\pi \kB T \tau)^{-1}$ is a number $<1$ for
dirty superconductors and $=1$ in the clean limit \cite{BCSGL}.
An immediate consequence of (\ref{ZI}) is the validity of the asserted
relation $\xi \gsim a$ at any temperature $T$.
$\xi/a$ diverges as $T$ approaches $\Tco$ from below; note that one usually
even assumes $\xi \gg a$ to justify a local GL description.
 From BCS theory we can also estimate the value of $\xi/a$ in the
interesting region close to the presumed vortex unbinding transition: using
(\ref{ZC}), (\ref{ZI}) one can show that
\be
  \frac{\xi}{a} \approx \sqrt{\frac{\hbar^2 \ntwoD}{2\mpar \kB T K}}
    \approx 210 \sqrt{\frac{\ntwoD[\Ang^{-2}]}{T[\Kelvin] \,
K}\frac{\me}{\mpar}} \ ,
\label{ZJ}
\ee
where $\ntwoD[\Ang^{-2}]$ is the number of carriers in the film per $\Ang^2$
and $\mpar$ their in-plane effective mass.
At the transition ($T = \Tv$), we assume again $K \approx 1$ and (\ref{ZJ})
yields a $\xi(\Tv)$ which is appreciably larger than $a$ unless the film has
at the same time a high $\Tco$, low carrier density $\ntwoD$ and high in-
plane carrier mass $\mpar$ (note that these conditions may be realizable in
films of high-$\Tc$ material whose thickness is a few unit cells).

The last point we want to briefly address in this subsection is the omission
of a vector potential term in (\ref{ZA}).
{\em A priori}, such a coupling is important also in the absence of an
external magnetic field, since thermal $\psi$ fluctuations are accompanied
by local supercurrents and therefore generate local magnetic fields.

The argument for the conventional scenario (no fluctuations except vortices)
is well known \cite{BMO79,HN79} and goes as follows.
It was shown by Pearl \cite{Pea64} that the logarithmic vortex-vortex
interaction in superconducting films is magnetically screened at a length
\be
  \Lams = \frac{2\lamL^2}{d} \ ,
\label{ZK}
\ee
where $\lamL$ is the bulk London penetration depth and $d$ the film
thickness.
For this reason, KT originally argued that a vortex unbinding transition
should not occur in superconducting films since the logarithmic interaction
at arbitrarily large distances is essential for the mechanism.
However, it was noticed later \cite{BMO79,HN79} that for small $d$ and
sufficiently close to $\Tco$ (note that $\lamL$ diverges as $T \to \Tco$),
$\Lams$ can become of the order of millimeters, i.e.\ comparable to typical
sample sizes.
Screening then should not wash out the transition to a larger extent than
effects of finite system size and is therefore disregarded.
The above authors restricted their argument to dirty films with a large
$\lamL(T=0)$ but one can easily convince oneself using the GL relationship
$\lamL^2 = \hbar^2 c^2 \beta/32\pi |\alpha| \gamma e^2$ that for {\em any}
film, independently of its specifications,
\be
  \Lams = \frac{(\hbar c/e)^2}{8\pi K \kB T} \approx \frac{1.2\,
\mbox{cm}}{T[\Kelvin] \, K} \ .
\label{ZL}
\ee
At $T = \Tv$, $\Lams$ should thus always be roughly of the order of 1 mm.

In this paper, we consider --- in addition to vortices --- the effect of
amplitude and phase modulations which are of course also accompanied by
supercurrents.
Screening effects lead in this case to an additional ``mass" term $\sim
2/\Lams^2$ for fluctuations, in particular the massless phase (Goldstone)
modes of the neutral superfluid become slightly massive.
However, since our calculations will involve fluctuations of wavelength
$\lsim \xi$ only and since we can assume $\xi \ll \Lams$ because of
(\ref{ZL}), we believe that we may disregard this effect, too.
A more detailed discussion of this assumption may be interesting, but we
will not further enter this question here.

\subsection{Elimination of short-wavelength fluctuations: mapping on the
two-dimensional Coulomb gas}

In this section, we outline the basic ideas of our approach.
In particular, we try to clarify how the mapping of the 2DGL model at finite
$T$ on its associated vortex gas can be carried out in a systematic way.

Our aim is to study the presumed vortex unbinding transition which should
show up in thermodynamic quantities like the partition sum (\ref{ZB}).
To this end we try to transform (\ref{ZB}) into the statistical-mechanical
partition sum of an interacting vortex gas, i.e.\ we try to eliminate any
degrees of freedom except the vortex positions.

On the technical level, the main step is to introduce a new lattice cutoff
$\ta > a$ and to define an effective action $\tS[\psi_0]$ on a coarser
lattice with constant $\ta$ (i.e., of the long-wavelength field components
$\psi_0$) by integrating out the short-wavelength components $\psi_1$ which
contain wave numbers between $\pi/\ta$ and $\pi/a$:
\be
  Z \propto \int_{0 < k < {\textstyle \frac{\pi}{\ta}}} {\cal D}\psi_0 \,
e^{\tS[\psi_0]} \ ,
\label{AD}
\ee
\be
  e^{\tS[\psi_0]} :=
  \int_{{\textstyle \frac{\pi}{\ta}} < k < {\textstyle \frac{\pi}{a}}}
  {\cal D}\psi_1 \, e^{-\frac{K}{2} H[\psi_0 + \psi_1]} \ .
\label{AE}
\ee
If $\ta$ is chosen large enough ($\ta \gsim a,\xi$) then we expect
(\ref{AD}) to be dominated by ``saddle point" configurations $\psi_0$ of $H$
with vortices at given positions $\vR_i$ (and with associated vorticities
$m_i$) as constraints.
The effective phase space of the resulting vortex gas will be formed by the
plaquettes (with area $\ta^2$) of the {\em coarser} lattice which is
important for the definition of the vortex entropy and thus for the
thermodynamics.
Non-singular phase fluctuations (``spin waves") of wave numbers $k <
\pi/\ta$ are also still contained in (\ref{AD}), but they decouple from the
vortices like they do in the pure XY model and play no role in the
transition, so we will disregard them (see also subsection 4.2).

On the other hand, if $\ta$ is not much larger than $\max(a,\xi)$, then the
effective action $\tS[\psi_0]$ may be determined to a reasonable
approximation by a gaussian approximation to the functional integral in
(\ref{AE}).
$\ta$ may then be thought of as the smallest length scale on which vortices
are well distinguished objects.
In subsection 4.2 we will realize the infrared cutoff $\pi/\ta$ by a mass
$2/\xi^2$ of the field $\psi_1$ (``soft" cutoff), where $\xi$ is the GL
correlation length.
Both quantities will then turn out to be related by $\ta^2 = a^2 + 2\pi
\xi^2$ which is in good agreement with our general discussion in the
introduction and satisfies the above criteria.

Disregarding phase fluctuations of wave numbers $k < \pi/\ta$ as stated
above,  the vortex part of (\ref{AD}) can be expressed as
\be
  Z_{\rm v} = \sum_N \frac{1}{N!} \sum_{\{m_i\}} \left( \prod_{i=1}^N \int
\frac{d^2 \vR_i}{\ta^2} \right) e^{\tS[\psi_0^{(N)}]} \ ,
\label{AF}
\ee
where the determination of $\tS[\psi_0^{(N)}]$ as a function of the
$\vR_i,m_i$ from (\ref{AE}) is the main technical task in this paper.
$N$ is the total number of vortices (of mutual separation  $\gsim \ta$) in
the system.
We will later see that thermodynamically (i.e., in the partition sum
(\ref{AF})) only vorticities $m_i = \pm 1$ are important and, moreover,
below $\Tv$ only ``neutral" vortex configurations (obeying $\sum_i m_i =
0$).

As we will see in subsection 4.3, for vortex distances much larger than
$\ta$, $\tS[\psi_0^{(N)}]$ has an asymptotic behavior of the form
\be
  \tS[\psi_0^{(N)}]
  \sim \frac{1}{\TCG} \left( \sum_{i<j}^N m_i m_j \ln \frac{|\vR_i -
\vR_j|}{\ta} + N \muCG \right) \ ,
\label{AG}
\ee
i.e.\ it behaves as a neutral two-dimensional Coulomb gas (2DCG).
The latter is generally defined by a partition sum of the form \cite{Min87}
\be
  Z_{\rm v} = \sum_N \frac{1}{N!} {\sum_{\{m_i\}}}' \left( \prod_{i=1}^N
\int \frac{d^2 \vR_i}{\Delta} \right)
\exp \frac{1}{\TCG} \left( \sum_{i<j}^N m_i m_j \ln_+ \frac{|\vR_i -
\vR_j|}{\Rc} + N \muCG \right) \ ,
\label{AH}
\ee
where $\sum'_{\{m_i\}}$ now is a restricted sum over all {\em neutral}
configurations of $N$ vortices with $m_i = \pm 1$.
$\Delta$ is the ``phase space division" of a CG charge, $\TCG$ the
dimensionless CG temperature and $-2\muCG$ may be interpreted as the
creation energy of a neutral pair at distance $\Rc$.
Note that in (\ref{AH}) the interaction involves $\ln_+ x := \max (0,\ln
x)$, i.e.\ it is zero at smaller distances than $\Rc$.
Such a cutoff is essential for (\ref{AH}) to be well-defined, but it can be
realized in different ways \cite{Min87}.
It is easy to see that despite the appearence of the four parameters $\TCG,\
\muCG,\ \Delta$ and $\Rc$, (\ref{AH}) actually depends only on $\TCG$ and on
the ``fugacity"
\be
  \zCG = \frac{\Rc^2}{\Delta} e^{\muCG/\TCG}
\label{AI}
\ee
of the CG charges.
The values which were used to draw the $\zCG(\TCG)$ lines in fig.\ \ref{CG1}
are taken from refs.\ \cite{JKK77} (XY model) and \cite{Min87,WM88} (``bare"
GLCG), and are collected in table I.
For the effective GLCG defined by (\ref{AF}), (\ref{AG}) we have by
definition $\Delta = \ta^2$, but the short-distance cutoff is not yet
specified in (\ref{AG}).
We assume here $\Rc \approx \ta$, such that simply $\zCG = e^{\muCG/\TCG}$.
This is reasonable since $\ta$ is of the order of the vortex core size.
To demonstrate that the results for $\zCG(\TCG)$ are not too sensitive to
the choice of $\Rc$ (nor presumably to the precise cutoff procedure), we
have changed the values of the cutoff distance $\Rc$ in the case of the XY
model and the ``bare" GLCG by factors of 2 and 1/2.
Note that there is an accompanying shift of $\muCG$: the change to some
other cutoff $\Rc'$ in (\ref{AH}) means that we have to replace $\ln_+
(R/\Rc) \mapsto \ln_+ (R/\Rc') + \ln (\Rc'/\Rc)$ which (using the charge
neutrality condition) leads to
\be
  z \mapsto z'= z \left( \frac{\Rc'}{\Rc} \right)^{2 - 1/2\TCG} \ .
\label{AJ}
\ee
The results are shown in fig.\ \ref{CG2} and present no essential changes
compared to fig.\ \ref{CG1}; the effects we will be concerned with later on
are much more pronounced.

\section{The ``saddle point" configuration}

As we argued in subsection 2.2, the long-wavelength component $\psi_0$ of
the field is essentially of the form of a ``saddle point" configuration of
$H$ for given vortex positions $\vR_i$ and vorticities $m_i$.
So our first step will be a detailed investigation of these configurations
and in particular of their energies $H[\psi_0]$.
In section 4 we will then proceed calculating the effective action
$\tS[\psi_0]$ of (\ref{AE}), by adding corrections due to thermal
fluctuations around these saddle points.

Since in this section we are not considering fluctuations, we can work in
the continuum limit $a \to 0$ (such that $\xi$ is the only length scale of
the hamiltonian (\ref{ZD})) without encountering divergencies.
Only in subsection 3.5 we will reintroduce a discrete lattice and discuss,
in which way our results then are modified.

\subsection{Separation of the vortex degrees of freedom}

In order to see how vortices, i.e.\ singularities of the phase with finite
winding numbers (``vorticities") can be imposed as constraints on the field
$\psi$, we write the latter in terms of real fields $\rho,\theta$
(``amplitude" and ``phase"),
\be
  \psi = \rho e^{i \theta}
\label{BA}
\ee
and express the hamiltonian (\ref{ZD}) in terms of $\rho,\theta$:
\be
  H[\rho,\theta] =
    \int d^2 \vr \left\{ \frac{1}{2 \xi^2}(1 - \rho^2)^2 + |\grad \rho|^2
    + \rho^2 |\grad \theta|^2 \right\} \ .
\label{BB}
\ee
The phase gradient can be split into its longitudinal and transverse parts,
\be
  \grad \theta = \grad \vartheta - \vn \times \grad \Phi \ ,
\label{BC}
\ee
where $\vn$ is the unit vector normal to the plane.
$\vartheta$ is a nonsingular phase field representing the ``spin waves", and
$\Phi$ is a ``vortex potential" which satisfies the Poisson equation
\be
  \grad^2 \Phi(\vr) = - 2\pi \sum_i m_i \delta(\vr - \vR_i)
\label{BD}
\ee
with point-like integer ``charges" (vorticities) $m_i$ at the positions
$\vR_i$, corresponding to the singularities of the phase field $\theta$.
($\theta$ changes by $2\pi m_i$ upon going counterclockwise around $\vR_i$).
$\Phi$ is determined by (\ref{BD}) only up to a harmonic function; any
harmonic contribution can however be absorbed into $\vartheta$, so that we
may choose the particular solution
\be
  \Phi(\vr) = - \sum_i m_i \ln \frac{|\vr - \vR_i|}{\xi}
\label{BE}
\ee
which finally renders the splitting (\ref{BC}) unique.
In order to make the argument of the logarithm in (\ref{BE}) dimensionless
we have introduced the only length scale $\xi$ of the problem.
Note however that for a ``neutral" vortex configuration $(\sum_i m_i = 0),\
\Phi$ does not depend on $\xi$.
The physical content of (\ref{BC}), (\ref{BE}) is that we have separated the
vortex degrees of freedom (expressed by their positions $\vR_i$ and
vorticities $m_i$) from the remaining nonsingular phase configuration
$\vartheta$.

\subsection{The vortex core structure (1-vortex problem)}

Besides the phase field which defines the vortices, our hamiltonian contains
the amplitude field $\rho$ which is strongly coupled to $\theta$ close to
the vortex centers $\vR_i$: far away from the centers ($\grad \theta$ small)
it is expected to have values $\rho \approx 1$, whereas it vanishes right at
the vortex centers, $\rho(\vR_i) = 0$, because otherwise the ``vortex core"
energies would diverge logarithmically in the continuum limit.
We call {\em vortex core} the regions of size $\approx \xi$ around $\vR_i$
where $\rho$ significantly differs from 1.
Since the structure of the core regions is essential for the following, we
will investigate here in detail a single isolated vortex of vorticity $m$
centered at the origin.
This subsection is sort of a summary of results taken from the literature
which are relevant in our context.

Because of the isotropy of the problem we employ polar coordinates $r,\phi$;
in the saddle-point configuration of the hamiltonian (\ref{BB}), the phase
field is then (up to an additive constant which we choose equal to zero)
given by
\be
  \theta_m(r,\phi) = m \phi
\label{CAa}
\ee
and the amplitude $\rho_m$ is a function of $r$ alone.
Inserted in (\ref{BB}), this yields a reduced hamiltonian of the 1-vortex
problem,
\be
  H_m[\rho_m] =
    \int dr \, 2\pi r \left\{
      \frac{1}{2 \xi^2}(1 - \rho_m^2)^2
      + \left( \frac{d \rho_m}{dr} \right)^2
      + \frac{m^2}{r^2} \rho_m^2
    \right\} \ .
\label{CA}
\ee
Its minimum solution is determined by a vanishing functional derivative with
respect to $\rho_m$,
\be
  0 = - \frac{1}{4 \pi r} \frac{\delta H_m}{\delta \rho_m(r)}
    = \frac{d^2 \rho_m}{dr^2} + \frac{1}{r} \frac{d \rho_m}{dr}
      - \frac{m^2}{r^2} \rho_m + \frac{1}{\xi^2} \rho_m (1 - \rho_m^2) \ ,
\label{CBa}
\ee
together with the boundary conditions
\be
  \rho_m(0) = 0 \ , \quad
  \rho_m(r) \to 1 \mbox{\quad for \quad} r \to  \infty \ .
\label{CBb}
\ee
The solution $\rho_m(r)$ rises monotonically from 0 at $r=0$ to 1 at $r \to
\infty$.
To determine the asymptotic behavior of $\rho_m$ at small and large $r$, one
may proceed as follows:

\begin{enumerate}

\item
$r \ll \xi$: here $\rho_m(r) \ll 1$, so we linearize (\ref{CBa}) in $\rho_m$
and insert an ansatz $\rho_m(r) \sim c_m(r/\xi)^\alpha$ with $\alpha > 0$
and $c_m$ a numerical constant, which yields
\be
  0 \sim \frac{d^2 \rho_m}{dr^2} + \frac{1}{r} \frac{d \rho_m}{dr}
         - \frac{m^2}{r^2} \rho_m
    \sim c_m (\alpha^2 - m^2) \frac{r^{\alpha - 2}}{\xi^\alpha} \ .
\label{CC}
\ee
Thus $\alpha = |m|$, and $c_m$ is a constant of order 1 which has to be
determined numerically from the full solution of (\ref{CBa}), (\ref{CBb}).
For $m = 1$, reference \cite{MN85} gives
\be
  \lim_{r \to 0}
  \left( \frac{\xi^2}{2\pi r} \frac{d}{dr} \rho_{m=1}^2 \right)
  = \frac{c_{m=1}^2}{\pi} = 0.108 \ ,
    \mbox{\quad i.e. \quad} c_{m=1} = 0.582 \ .
\label{CD}
\ee

\item
$r \gg \xi$: here $u(r) := 1-\rho_m(r) \ll 1$, so we linearize (\ref{CBa})
in $u$ and insert an ansatz $u(r) \sim c'_m(r/\xi)^\beta$ with $\beta < 0$:
\bea
  0 & \sim & \frac{d^2 u}{dr^2} + \frac{1}{r} \frac{du}{dr}
         + \frac{m^2}{r^2} (1 - u) - \frac{2}{\xi^2} u
\nonumber \\
    & \sim & c'_m(\beta^2 - m^2) \frac{r^{\beta - 2}}{\xi^\beta}
         + \frac{m^2}{r^2} - 2 c'_m \frac{r^\beta}{\xi^{\beta+2}} \ .
\label{CE}
\eea
For $r \to \infty$, the last two terms are the dominant ones, such that
$\beta = -2$ and $c'_m = m^2/2$.

\end{enumerate}
In conclusion, the asymptotic behavior of $\rho_m$ is given by
\be
  \rho_m(r) \sim \left\{
    \begin{array}{ccc}
      c_m (r/\xi)^{|m|} & \mbox{for} & r \ll \xi \\
      {\displaystyle 1 - \frac{m^2}{2} (\xi/r)^2 } & \mbox{for} & r \gg \xi
    \end{array}
  \right. \ , \quad
  c_{m=1} = 0.582 \ .
\label{CF}
\ee
Sketches of $\rho_m$ for $m = 1,2$ are given in reference \cite{MN85}.
For later use we note that the first line of (\ref{CF}) together with
$\theta_m = m \phi$ imply that close to the vortex center ($r \ll \xi$) the
field $\phi_m = \rho_m e^{i \theta_m}$ has the simple power form
\be
  \psi_m(\hr) \sim \left\{
    \begin{array}{ccc}
      c_m (\hr/\xi)^{|m|}   & \mbox{for} & m > 0 \\
      c_m (\hr^*/\xi)^{|m|} & \mbox{for} & m < 0
    \end{array}
  \right. \ ,
\label{CFa}
\ee
where $\hr = r e^{i \phi}$ is a complex notation for the coordinate in the
plane.
Note in particular that the complex field is perfectly smooth even in the
center of a vortex, a singularity only appears when one looks at the phase
separately.

Now consider the different contributions to the total energy $H_m$
(\ref{CA}).
Since for a circle-shaped system $H_m$ diverges with the radius $r_c$ as
$\sim 2\pi m^2 \ln r_c$, we split off this size-dependent term:
\be
  H_m[\rho_m] \sim 2\pi \left( m^2 E_1(m) + E_2(m) + \frac{E_3(m)}{2}
    + m^2 \ln \frac{r_c}{\xi} \right)
\label{CG}
\ee
for $r_c \gg \xi$, where
\bea
  E_1(m) & := & \lim_{r_c \to \infty}
    \left( \int_0^{r_c} dr \frac{\rho_m^2}{r} - \ln \frac{r_c}{\xi} \right)
    = - \int_0^\infty dr \frac{d \rho_m^2}{dr} \ln \frac{r}{\xi} \ ,
\label{CHa} \\
  E_2(m) & := & \int_0^\infty dr\, r \left( \frac{d \rho_m}{dr} \right)^2 \
,
\label{CHb} \\
  E_3(m) & := & \int_0^\infty dr \frac{r}{\xi^2} (1 - \rho_m^2)^2 \ .
\label{CHc}
\eea
The second expression for $E_1(m)$ follows by partial integration.

$E_3(m)$ can be calculated analytically by the following trick (we follow an
idea of reference \cite{MN85}):
let $\rho_m$ be the solution of (\ref{CBa}), (\ref{CBb}) and let
$\rho_{m,\alpha}(r) := \rho_m(r/\alpha)$.
Then the energy $H_m[\rho_{m,\alpha}]$ is minimum for $\alpha = 1$.
To yield finite results it must again be regularized by a finite system
radius $r_c \gg \xi$.
After a variable change $r \mapsto \alpha r$ it reads:
\be
  H_m[\rho_{m,\alpha}]  =
    \int_0^{\alpha r_c} dr 2\pi r \left\{
      \frac{\alpha^2}{2 \xi^2}(1 - \rho_m^2)^2
      + \left( \frac{d \rho_m}{dr} \right)^2
      + \frac{m^2}{r^2} \rho_m^2
    \right\} \ .
\label{CI}
\ee
The minimum condition then implies that for $r_c \gg \xi$:
\be
  0 = \frac{1}{2\pi} \left. \frac{d H_m}{d \alpha} \right|_{\alpha = 1}
  \sim m^2 \rho_m(r_c)^2 - \int_0^{r_c} dr \frac{r}{\xi^2} (1 - \rho_m^2)^2
  \sim m^2 - E_3(m) \ ,
\label{CJa}
\ee
i.e.
\be
  E_3(m) = m^2 \ .
\label{CJ}
\ee
Once (\ref{CBa}), (\ref{CBb}) is numerically solved for a given $m$, the
numbers $E_1(m)$ and $E_2(m)$ can also be evaluated.
Ref.\ \cite{Hu72} supplies a value $E_2(1) = 0.279$.
This value is confirmed by ref.\ \cite{MN85} which furthermore supplies
$E_2(2)/4 + E_3(2)/8 = 0.604$, i.e.\ $E_2(2) = 0.416$ and $E_1(1) = -\ln
2.24 = -0.806$.
The mentioned values of $E_i(m)$ are summarized in table II.

\subsection{Interaction of vortices at large distances}

Let us now calculate the interaction energy of a given vortex configuration
defined by its potential $\Phi$ (\ref{BE}), taking the core structure of the
vortices into account.
The interesting term in the hamiltonian (\ref{BB}) is the third one which
couples $\rho$ and $\theta$.
Inserting (\ref{BC}), we obtain
\bea
  \int d^2 \vr \, \rho^2 |\grad \theta|^2
  & = & \int d^2 \vr \, \rho^2 |\grad \vartheta
    - \vn \times \grad \Phi|^2
  \nonumber \\
  & = & \int d^2 \vr \, \rho^2 \left\{ |\grad \vartheta|^2
    - 2 \grad \vartheta \cdot (\vn \times \grad \Phi)
    + |\grad \Phi|^2 \right\}
  \label{DA} \\
  & = & \int d^2 \vr \left\{ \rho^2 |\grad \vartheta|^2
    - 2 \vartheta \vn \cdot (\grad \rho^2 \times \grad \Phi)
    - \Phi (\rho^2 \grad^2 \Phi
    + \grad \rho^2 \cdot \grad \Phi) \right\} \ .
  \nonumber
\eea
The boundary terms appearing in the partial integration which leads to the
third line of (\ref{DA}) would yield a {\em positive} contribution to the
total energy which diverges logarithmically with the system size unless the
vortex configuration is neutral ($\sum_i m_i = 0$), in which case they
vanish.
Non-neutral configurations are therefore thermodynamically suppressed (at
least at low temperatures), and we consider here only neutral ones, for
which (\ref{DA}) is correct.

To find the saddle point field configuration for given vortices we now have
to minimize (\ref{BB}) (with (\ref{DA}) inserted) with respect to $\rho$ and
$\vartheta$.
Because of (\ref{BD}) and $\rho(\vR_i) = 0$ the $\rho^2 \grad^2 \Phi$ term
vanishes identically.
In general the other terms interact in a complicated manner. However, we are
interested at vortices at distances $\gsim \xi$ since we will treat the
short-range fluctuations by other means in section 4.
Let us for simplicity assume that the smallest distance $|\vR_i - \vR_j|$
between any two vortices in the given configuration is $R \gg \xi$ and look
for the leading terms in an expansion in $\xi/R$.
The vortex cores are now well separated and in any of them $\rho$ is
expected to be given by the isotropic 1-vortex solution of subsection 3.2
plus a correction of the order $\order(\xi/R)$ (a justification will be
given below).
Together with (\ref{BE}) and by symmetry arguments this implies that $\grad
\rho^2 \times \grad \Phi$ is of order $\order(\xi/R)$ which in turn gives
(see (\ref{DA})) a saddle point configuration of $\vartheta$ of order
$\order(\xi/R)$.
(\ref{DA}) therefore yields
\bea
  \lefteqn{
    \int d^2 \vr \rho^2 |\grad \theta|^2 =
    - \int d^2 \vr \, \Phi (\grad \Phi \cdot \grad \rho^2) +
\order(\xi^2/R^2)
  } \nonumber \\
  & & = - \sum_{i,j} m_i m_j \int d^2 \vr
      \frac{(\vr - \vR_j) \cdot \grad \rho^2}{|\vr - \vR_j|^2} \ln
\frac{|\vr - \vR_i|}{\xi}
      + \order(\xi^2/R^2)
  \ . \label{DB}
\eea
In the second line, simply (\ref{BE}) has been inserted.
The important contributions to the integral in (\ref{DB}) stem from a region
of size $\order(\xi)$ around $\vR_j$, the rest is again of order
$\order(\xi^2/R^2)$.
To calculate this integral, we may then assume that $\vR_j = 0$ and replace
$\rho$ by the 1-vortex solution with vorticity $m_j$ centered at the origin,
$\rho_{m_j}$ (see subsection 3.2).
We further have to distinguish two cases:

\begin{enumerate}

\item
$i \neq j$; i.e., $\vR_i = \vR_i - \vR_j =: \Delta\vR,\ |\Delta\vR| \geq R$.
In this case the logarithm in (\ref{DB}) is dominated by a constant term
$\ln(|\Delta\vR|/\xi)$ and contains further terms of order $\order(\xi/R)$
which couple $\rho$ in the core of vortex $j$ to the other vortices.
This leads to an {\em a posteriori} justification of the above assumption
that the corrections to the 1-vortex solution $\rho_{m_j}$ around $\vR_j$
are of the order $\order(\xi/R)$.
Again invoking symmetry arguments we finally obtain
\bea
  \int d^2 \vr \frac{\vr \cdot \grad \rho^2}{|\vr|^2} \ln \frac{|\vr -
\Delta \vR|}{\xi}
  & = & 2\pi \ln \frac{|\Delta \vR|}{\xi} \int_0^\infty dr
        \frac{d \rho_{m_j}^2}{dr}
        + \order(\xi^2/R^2)
  \nonumber \\
  & = & 2\pi \ln \frac{|\Delta \vR|}{\xi}
        + \order(\xi^2/R^2)
  \ , \label{DC}
\eea
since $\rho_{m_j}(r) \to 1$ for $r \to \infty$.

\item
$i=j$; i.e., $\vR_i = \vR_j = \vo$.
This gives the strong ``self-interaction" of the core amplitude of a vortex
with its own phase configuration,
\bea
  \int d^2 \vr \frac{\vr \cdot \grad \rho^2}{|\vr|^2} \ln \frac{|\vr|}{\xi}
\,
  & = & 2\pi \int_0^\infty dr \frac{d \rho_m^2}{dr} \ln \frac{r}{\xi}
        + \order(\xi^2/R^2)
  \nonumber \\
  & = & - 2\pi E_1(m)
        + \order(\xi^2/R^2)
  \ . \label{DD}
\eea

\end{enumerate}
We can now insert these results into (\ref{DB}), add the terms of (\ref{BB})
which contain $\rho$ only and express everything in terms of 1-vortex
quantities.
So if $\psi_0 = \psi_0(\{\vR_i,m_i\})$ is the minimum-energy field
configuration with given vortices of vorticities $m_i$ at positions $\vR_i$,
its energy is given by
\be
  H[\psi_0] =
    - 4\pi \sum_{i<j} m_i m_j \ln \frac{|\vR_i - \vR_j|}{\xi}
    - 4\pi \sum_i \muGL(m_i)
    + \order(\xi^2/R^2) \ ,
\label{DE}
\ee
where
\be
  - 2 \muGL(m) = m^2 E_1(m) + E_2(m) + \frac{m^2}{2} \ .
\label{DEa}
\ee
To 0th order in $\xi/R$, this is just the hamiltonian of a {\em neutral two-
dimensional Coulomb gas} of integer charges $m_i$ whose core energy (or
chemical potential $\muGL$) depend on $m_i$.
Note that in particular (\ref{DE}) contains {\em two-body} interactions
only.
We will later absorb the overall factor $4\pi$ in the effective Coulomb gas
temperature.

For later use we also calculate the quantity $\xi^{-2} \int d^2 \vr (1 -
|\psi_0|^2)$ by relating it to $H[\psi_0]$ in the following way: since
$\psi_0$ is the minimum-energy field configuration for a given vortex
configuration $\{\vR_i,m_i\}$ as boundary conditions, the functional
derivative $\delta H / \delta \psi_0(\vr)$ vanishes nowhere in space except
at the vortex centers $\vr = \vR_i$.
On the other hand, $\psi_0(\vr)$ itself has zeros precisely at the vortex
centers, such that (with $\psi_0 = \rho e^{i\theta}$):
\be
  0 = \int d^2 \vr \, \psi_0 \frac{\delta H}{\delta \psi_0} =
    \int d^2 \vr \left\{ - \frac{1}{\xi^2} \rho^2 (1 - \rho^2) + |\grad
\rho|^2
    + \rho^2 |\grad \theta|^2 \right\} \ .
\label{DG}
\ee
Now after adding $\xi^{-2} \int d^2 \vr (1 - |\psi_0|^2)$ on both sides, the
r.h.s.\ becomes almost identical to $H[\psi_0]$ apart from an additional
factor of 2 in front of the $E_3$ integral (see (\ref{CHc})).
In analogy to (\ref{DE}) we therefore obtain
\bea
  \int \frac{d^2 \vr}{\xi^2} (1 - |\psi_0|^2) & = &
    - 4\pi \sum_{i<j} m_i m_j \ln \frac{|\vR_i - \vR_j|}{\xi}
    - 4\pi \sum_i \left( \muGL(m_i) - \frac{1}{4} m_i^2 \right)
    + \order(\xi^2/R^2)
  \nonumber \\
  & = & H[\psi_0] + \pi \sum_i m_i^2 + \order(\xi^2/R^2)
  \ . \label{DH}
\eea
Identity (\ref{DH}) will be of central importance in section 4 for the
mapping of the full problem including fluctuations on a Coulomb gas.

At low temperatures $T < \Tv$, the partition sum (\ref{AF}) will be
dominated by configurations made out of bound vortex pairs of opposite
vorticity.
Furthermore, higher vorticities $|m| \geq 2$ have a much lower statistical
weight than $|m| = 1$ for pair distances $R \gg \xi$ because of their higher
core energy and stronger binding (see (\ref{DE})).
We therefore assume that they play no essential role in the vortex unbinding
transition and restrict ourselves in the following to configurations with
$m_i = \pm 1$.

\subsection{The functional derivative $\delta H / \delta \psi_0$}

The field configuration $\psi_0$ which we studied in subsection 3.3 is no
real saddle point of $H$ since it is subject to the vortex boundary
conditions defined by $\{\vR_i , m_i\}$.
However, clearly $\delta H / \delta \psi_0(\vr_i) = 0$ for $\vr \not\in
\{\vR_i\}$, so we expect something like $\delta H / \delta \psi_0(\vr_i)
\propto \delta(\vr - \vR_i)$ for $\vr$ in the vicinity of $\vR_i$.
Since the functional derivative $\delta H / \delta \psi_0$ determines the
{\em linear} energy of fluctuations away from $\psi_0$ (see section 4), we
will study it here in some detail.
We consider again a neutral vortex configuration with minimum vortex
distance $R \gg \xi$.
Let us look at a neighborhood of the position $\vR_i$ of the $i$th vortex
and consider the particular fluctuation $\delta \psi_0$ generated by moving
vortex $i$ by an infinitesimal vector $\delta \vR_i$.
(\ref{CFa}) then implies that for $|\vr - \vR_i| \ll \xi$ (assume for
simplicity that $m > 0$),
\be
  \psi_0(\vr) \sim
    c_m \left( \frac{\hr - \hR_i}{\xi} \right)^m
    + \order(\xi/R)
\label{FA}
\ee
with complex representations $\hr,\hR_i$ for $\vr,\vR_i$ as in (\ref{CFa}).
Consequently,
\be
  \delta \psi_0(\vr) \sim
    - \frac{c_m}{\xi^m} \left( \hr - \hR_i \right)^{m-1} \delta \hR_i
    + \order(\xi/R) \ .
\label{FB}
\ee
Now (\ref{DE}) implies that up to terms of order $\order(\xi^2/R^2)$ the
total energy of the vortex system changes by an amount
\be
  \delta H[\psi_0]
    = - \vF_i \cdot \delta \vR_i
    = - \half \left( \hF_i^* \delta \hR_i + \hF_i \delta \hR_i^* \right) \ ,
\label{FC}
\ee
where
\be
  \vF_i :=
    4\pi m_i \sum_{j(\neq i)} m_j \frac{\vR_i - \vR_j}{|\vR_i - \vR_j|^2}
\label{FD}
\ee
is the total Coulomb force on the $i$th vortex, which is of the oder
$\order(1/R)$.
On the other hand, we can of course write
\be
  \delta H[\psi_0] =
    \int d^2 \vr \left( \frac{\delta H}{\delta \psi_0} \delta \psi_0
    + \frac{\delta H}{\delta \psi_0^*} \delta \psi_0^* \right) \ ,
\label{FE}
\ee
which after insertion of (\ref{FB}) leads by comparison with (\ref{FC}) to
\be
  \left( \frac{\hr - \hR_i}{\xi} \right)^{m-1}
  \frac{\delta H}{\delta \psi_0(\vr)}
    = \frac{\xi}{2 c_m} \hF_i^* \delta(\vr - \vR_i) + \order(\xi^2/R^2)
\label{FF}
\ee
Note that the factor in front of the $\delta$ function is of the order
$\order(\xi/R)$, which will permit us to neglect the corresponding terms in
the perturbation expansion of section 4.

\subsection{Interpolation to the XY model}

For convenience, we worked from the beginning of section 3 in the continuum
limit $a/\xi \to 0$.
However, when we include fluctuations around the saddle point configuration
$\psi_0$ in section 4, we will have to impose a finite lattice cutoff $a$.
We will then want to expand around a suitable saddle point for {\em finite}
$a/\xi$.
For the fluctuation corrections of section 4 we will take the discreteness
into account mainly by a suitable cutoff in momentum space.
The question addressed in this subsection is, what are the necessary
modifications of (\ref{DE}) respectively (\ref{DH}) for finite $a/\xi$?

The asymptotic behavior of the vortex potential at large distances $R \gg
\xi,a$ will not be affected by the discreteness of the system.
However, one expects an appreciable effect on vortex core energies since in
the core regions the field $\psi$ always varies on length scales down to the
order of the lattice constant $a$.
The extremest example is the XY or ``plane rotator" model, which corresponds
to the $\xi/a \to 0$ limit of (\ref{ZF}), (\ref{ZD}) where $\rho = |\psi|
\equiv 1$.
XY vortices are known to interact at distances $\gg a$ by a potential $4\pi
\ln(R/a)$ and to have a well-defined {\em finite} core energy $4\pi \muXY$
(which is approximately given by half the creation energy of a vortex pair
centered at neighboring plaquettes of the lattice), whereas (\ref{DE}) would
predict a {\em logarithmic divergency} of $\muXY$ with $\xi/a$.
More precisely, the energy calculated from (\ref{ZF}) respectively
(\ref{ZD}) of a neutral pair at distance $R \gg a,\xi$ has the asymptotic
forms
\be
  V_\xi(r) \sim \left\{
    \begin{array}{ccc}
    4\pi \left( \ln {\displaystyle \frac{R}{a}} - 2\muXY \right)   &
\mbox{for} & \xi \ll a \\
    4\pi \left( \ln {\displaystyle \frac{R}{\xi}} - 2\muGL \right) &
\mbox{for} & \xi \gg a \\
    \end{array}
  \right. \ ,
\label{EA}
\ee
where the respective chemical potentials are given by
\bea
  - 2 \muXY & = & \gamma + \frac{\ln 8}{2} \approx 1.617
                  \mbox{\quad (see ref.\ \cite{JKK77}),}
\nonumber \\
  - 2 \muGL & = & E_1(1) + E_2(1) + \half \approx -0.027 \ .
\label{EB}
\eea
Here $\gamma = 0.5772$ is the Euler constant and $E_1(1),\ E_2(1)$ are given
in table II.
The factor in front of the logarithm in (\ref{EA}) is not affected by the
lattice regularization.

We now look for a suitable interpolation formula for $V_\xi(R)$ which
correctly reproduces both limits (\ref{EA}).
Since we require an asymptotic behavior $V_\xi(R) \sim 4\pi \ln R + \const$
for all $\xi$, it should be of the form
\be
  V_\xi(r) \sim
    4\pi \left\{ \ln \frac{R}{a} - 2\muXY - F(\xi^2/a^2) \right\} \ ,
\label{EC}
\ee
where $F$ is an interpolating function with the asymptotic behavior
\be
  F(0) = 0 \ , \quad
  F(X) \sim \half \ln X + 2(\muGL - \muXY) \mbox{\quad for \quad} X \gg 1 \
{}.
\label{ED}
\ee
A simple realization of conditions (\ref{ED}) is
\be
  F(X) := \half \ln \left( 1 + X e^{4(\muGL - \muXY)} \right)
    \approx \half \ln \left( 1 + 26.78 X \right) \ .
\label{EE}
\ee
We tried to improve this simple guess for the interpolating function $F(X)$
by fitting the first derivative $F'(0)$ to the ``true" value estimated by
other means.
However, this kind of improvement had no substantial effect on our final
results, so we disregard it here.
Finally, using expressions (\ref{EC}), (\ref{EE}), the ``saddle point" field
energy (\ref{DE}) can be rewritten for vorticities $m_i = \pm 1$ in the
interpolated form (for $R \gg \xi,a$)
\be
  H[\psi_0] \approx
    - 4\pi \sum_{i<j} m_i m_j \ln \frac{|\vR_i - \vR_j|}{a}
    - 2\pi N (2 \muXY + F(X))
    + \order(\ta^2/R^2) \ ,
\label{EM}
\ee
where $N$ is the total number of vortices and $X = \xi^2/a^2$.

Starting from the lattice analog of (\ref{DG}) one can work out an
equivalent interpolation formula for the integral (\ref{DH}), which however
leads only to negligible corrections.
We will therefore later simply use the expression (\ref{DH}) with the
interpolated $H[\psi_0]$ inserted.

\section{Thermal fluctuations around the ``saddle point"}

As we already argued in the introduction we will identify the ``saddle
point" field configuration $\psi_0$ with the long-wavelength components of
the field $\psi$ with wave numbers $|\vk| < \pi/\ta$.
Note that the latter still contain long-wavelength phase fluctuations which
are absent in the ``saddle point" field configuration.
However, as we will argue in subsection 4.2 they are not important for the
transition and can be disregarded.
We will now calculate the effective action $\tS[\psi_0]$ (equations
(\ref{AD}), (\ref{AE})) in a Gaussian approximation, i.e.\ we expand
$H[\psi_0 + \psi_1]$ with respect to the fluctuations $\psi_1$ up to second
order and treat their coupling to the vortices perturbatively.

\subsection{Perturbational treatment of fluctuations}

The first step is to expand the GL functional (\ref{ZD}) around $\psi_0$:
\bea
  H[\psi_0 + \psi_1] & = & \int d^2 \vr \left\{
    \frac{1}{2\xi^2} \left( 1 - |\psi_0 + \psi_1|^2
    \right)^2
    + |\grad \psi_0 + \grad \psi_1|^2
    \right\}
  \nonumber \\
  & = & H[\psi_0] + \int d^2 \vr \left\{
    \frac{\delta H}{\delta \psi_0} \psi_1
    + \frac{\delta H}{\delta \psi_0^*} \psi_1^*
    \right\}
  \nonumber \\
  & & \quad + \int d^2 \vr |\grad \psi_1|^2
    + \int \frac{d^2 \vr}{2\xi^2} \left\{
    2 (2 |\psi_0|^2 - 1) |\psi_1|^2
    + \psi_0^{*2} \psi_1^2 + \psi_0^2 \psi_1^{*2}
    \right\}
  \nonumber \\
  & & \quad + \int \frac{d^2 \vr}{2\xi^2} \left\{
    2 (\psi_0^* \psi_1 + \psi_0 \psi_1^*) |\psi_1|^2
    + |\psi_1|^4
    \right\}
  \label{GA}
\eea
where the terms are ordered in ascending order in $\psi_1$.
In a vortex-free configuration we would have $|\psi_0| \equiv 1$ and the
linear terms in $\psi_1$ would vanish.
Now write again $\psi_0 = \rho_0 e^{i \theta_0}$, take all terms which are
quadratic in $\psi_1$ (and $\psi_1^*$), replace $\rho_0$ by 1 and collect
them in some ``free" fluctuation hamiltonian $H_0$.
(\ref{GA}) can now be rewritten as
\be
  H[\psi_0 + \psi_1] = H[\psi_0] + H_0[\psi_1] + H_I[\rho_0,\theta_0,\psi_1]
\label{GB}
\ee
where
\be
  H_0[\psi_1] = \int d^2 \vr \left\{
    \frac{1}{2\xi^2} \left( 2 |\psi_1|^2
      + \left( e^{-i\theta_0} \psi_1 \right)^2
      + \left( e^{i\theta_0} \psi_1^* \right)^2
      \right)
    + |\grad \psi_1|^2
    \right\}
\label{GC}
\ee
is the free fluctuation hamiltonian and all remaining terms are collected in
the ``perturbation"
\bea
  H_I[\rho_0,\theta_0,\psi_1] & = & \int d^2 \vr \left\{
    \frac{\delta H}{\delta \psi_0} \psi_1
    + \frac{\delta H}{\delta \psi_0^*} \psi_1^*
    \right\}
  \label{GD} \\
  & & \quad - \int \frac{d^2 \vr}{2\xi^2} (1 - \rho_0^2) \left\{
    4 |\psi_1|^2
    + \left( e^{-i\theta_0} \psi_1 \right)^2
    + \left( e^{i\theta_0} \psi_1^* \right)^2
    \right\}
  + \order(\psi_1^3)
  \nonumber
\eea
which describes the coupling of fluctuations to vortex cores and vanishes in
a vortex-free configuration.
Closer examination of (\ref{GC}) reveals that, locally, the field $\psi$ has
one massive and one massless component, the orientation of the local
reference frame being determined by the phase $\theta_0$ of the ``saddle
point" configuration.
More precisely, the real and imaginary parts of the ``gauge-transformed"
field $e^{- i \theta_0} \psi_1$ correspond to massive amplitude fluctuations
(mass $2/\xi^2$) and massless phase fluctuations, respectively.
In 2 dimensions the real-space propagator of a massless field is infrared
divergent, which causes serious problems in a perturbation expansion.
In the next section, we will therefore be concerned with the infrared
regularization of the phase fluctuations.

\subsection{Infrared regularization of phase fluctuations}

In order to understand the physical role of long-wavelength phase
fluctuations, we return to the representation (\ref{BB}), (\ref{DA}) of $H$,
where phase fluctuations are described by a field $\vartheta$.
The terms which couple to $\vartheta$ in (\ref{DA}) only contain wavelengths
$\lsim \xi$, so long-wavelength components $\vartheta_\vk$ with $|\vk| \ll
1/\xi$ decouple from the vortices and therefore play no role in the vortex
unbinding transition (note that in the continuum XY limit $\xi \ll a$,
$\vartheta$ decouples {\em completely!}).
This leads us to assume that all components with $|\vk| < \pi/\ta$ are
irrelevant and can be disregarded.
We will then treat the components with $\pi/\ta < |\vk| < \pi/a$
perturbatively.
For technical reasons it is convenient to realize this infrared momentum
cutoff $\pi/\ta$ in a ``soft" way, by assigning a {\em mass} to the phase
fluctuations in (\ref{GC}).
For simplicity we choose this mass to be the same ($= 2/\xi^2$) as for the
amplitude fluctuations, such that (\ref{GC}) is replaced by the simpler,
translationally invariant expression with isotropic mass matrix
\be
  \tH_0[\psi_1] = \int d^2 \vr \left(
    \frac{2}{\xi^2} |\psi_1|^2 + |\grad \psi_1|^2
    \right)
\label{HA}
\ee
(We will argue in subsection 4.4 that the relation between ``hard" and
``soft" cutoffs is given by $\ta^2 = a^2 + 2\pi \xi^2$).
With this modification, the hamiltonian (\ref{GB}) now describes a system of
interacting vortices together with a massive, harmonic field $\psi_1$ (made
up of amplitude and phase fluctuations and described by (\ref{HA})) which is
scattered by the vortex cores.
In the calculation of the effective action (\ref{AE}) we may now integrate
over {\em all} components $\psi_{1\vk}$ with $|\vk| < \pi/a$ since the
infrared cutoff is taken into account by the mass.

The free propagator corresponding to hamiltonian (\ref{HA}) is given by
\be
  \langle \psi_1(\vr) \psi_1^*(\vo) \rangle
    = \frac{ \int d[\psi_1] \psi_1(\vr) \psi_1^*(\vo) e^{- \tH_0} }
           { \int d[\psi_1] e^{- \tH_0} }
    = \frac{2}{K} G_0(\vr) \ ,
\label{HB}
\ee
where
\be
  G_0(\vr) = \int \frac{d^2 \vk}{(2\pi)^2} G_0(\vk) e^{i \vk \cdot \vr} \ ,
\label{HBa}
\ee
\be
  G_0(\vk) = \frac{1}{2/\xi^2 + \vk^2} \ .
\label{HBb}
\ee
For later use in the perturbational expansion, we derive here some
properties of $G_0$.
The local mean square value of $\psi_1$ is closely related to the
fluctuations already calculated in eq.\ (\ref{ZH}),
\be
  \frac{K}{2} \langle |\psi_1^2| \rangle = G_0(\vr = \vo)
  = \int \frac{d^2 \vk}{(2\pi)^2} \frac{1}{2/\xi^2 + \vk^2}
  \approx \frac{1}{4\pi} \ln(1 + 2\pi X) \ ,
\label{HBc}
\ee
where we again used the notation $X = \xi^2/a^2$.
More generally, one can show that
\be
  G_0(\vr) \approx \left\{
    \begin{array}{ccc}
    {\displaystyle \frac{1}{4\pi} \ln(1 + 2\pi X) } & \mbox{for} & \vr = \vo
\\
    {\displaystyle -\frac{1}{2\pi} \ln \frac{|\vr|}{\xi} } & \mbox{for} &
a\leq |\vr| \leq \xi \\
    0 & \mbox{for} & |\vr| > a,\xi \\
    \end{array}
  \right. \ ,
\label{HC}
\ee
where the second line is relevant only if $a < \xi$.
The $n$-fold convolution of $G_0$ with equal arguments at the ends can be
calculated in analogy to (\ref{HBc}),
\bea
  \lefteqn{ \left( G_0^{n+1} \right) (\vr = \vo)
  = \int_2 \cdots \int_{n+1} G_0(1,2) G_0(2,3) \cdots G_0(n+1,1)
  } \nonumber \\
  & = & \int \frac{d^2 \vk}{(2\pi)^2} G_0(\vk)^{n+1}
  \approx \frac{1}{4\pi} \int _0^{4\pi/a^2}
    \frac{d(k^2)}{(2/\xi^2 + k^2)^{n+1}}
  \nonumber \\
  & = & \frac{1}{4\pi n} \left\{
      \left( \frac{2}{\xi^2} \right)^{-n}
      - \left( \frac{4\pi}{a^2} + \frac{2}{\xi^2} \right)^{-n}
    \right\} \ ,
  \label{HD}
\eea
which is never larger than its $\xi \gg a$ limit:
\be
  \left( G_0^{n+1} \right) (\vr = \vo) \leq \frac{\xi^{2n}}{4\pi n 2^n} \ .
\label{HDa}
\ee
We write out the particular case $n=1$ explicitly:
\be
  \int d^2 \vr \, G_0(\vr)^2 \approx \frac{\xi^2}{4} \frac{X}{1 + 2\pi X}
  \leq \frac{\xi^2}{8\pi} \ .
\label{HE}
\ee

\subsection{Diagrammatics}

We will now calculate the effective action $\tS[\psi_0]$ (eq.\ (\ref{AE}))
by diagrammatic perturbation theory, starting from the decomposition
(\ref{GB}), (\ref{GD}), (\ref{HA}) of the original hamiltonian.
The propagator of the theory is $\frac{K}{2} G_0$, defined by (\ref{HBa}),
(\ref{HBb}).
Furthermore, up to order $\order(\psi_1^2)$ the interaction term $-
\frac{K}{2} H_I$ is represented by a sum of the diagram elements shown in
fig.\ \ref{diag1}, with combinatoric factors of 1/2 for the symmetric
diagrams (d) and (e).
The corresponding analytic expressions read
\bea
  \mbox{(a)} & = & - \frac{K}{2} \int \frac{\delta H}{\delta \psi_0^*}
\psi_1^*
\nonumber \\
  \mbox{(b)} & = & - \frac{K}{2} \int \frac{\delta H}{\delta \psi_0} \psi_1
\nonumber \\
  \mbox{(c)} & = & \frac{K}{\xi^2} \int (1 - \rho_0^2) |\psi_1|^2
\nonumber \\
  \mbox{(d)} & = & \frac{K}{2\xi^2} \int (1 - \rho_0^2) e^{2i\theta_0}
\psi_1^{*2}
\label{IA} \\
  \mbox{(e)} & = & \frac{K}{2\xi^2} \int (1 - \rho_0^2) e^{-2i\theta_0}
\psi_1^2 \ ,
\nonumber
\eea
respectively.
Omission of terms of higher order than $\order(\psi_1^2)$ just means that we
are working in the gaussian approximation.
The perturbation series for the effective action $\tS[\psi_0]$ (eq.\
(\ref{AE})) may now be expressed as $- \frac{K}{2} H[\psi_0]$ plus the sum
of the diagrams shown in fig.\ \ref{diag2} with the appropriate
combinatorial factors (plus higher order terms to be discussed in a moment).
We calculate these different diagrams separately, keeping only the
``leading" terms, i.e.\ those which contribute to the logarithmic
interaction or the chemical potential in $H[\psi_0]$ (eq.\ (\ref{DE})), and
omitting terms of order $\order(\xi^2/R^2)$.
\newline
Consider first the ``chain" diagrams in fig.\ \ref{diag2}:
\bea
  \mbox{(a)} & = & \frac{K}{2} \int_1 \int_2 G_0(1,2)
    \frac{\delta H}{\delta \psi_0(1)}
    \frac{\delta H}{\delta \psi_0^*(2)}
  \nonumber \\
  & \approx & \frac{K}{2} G_0(\vr = \vo) \left( \frac{\xi}{2 c_1} \right)^2
    \sum_i |\vF_i|^2 = \order(\xi^2/R^2) \ ,
  \label{IG}
\eea
where we used (\ref{FF}) for $m=1$.
Similar arguments show that all of the ``chain" diagrams ((b), (c), (d) and
``longer" ones) are likewise of order $\order(\xi^2/R^2)$ and they will
therefore be neglected.

The ``loop" diagrams (e), (f), ... in fig.\ \ref{diag2} are more
interesting.
The only one which gives a contribution to the logarithmic term in the
vortex interaction is diagram (e):
\bea
  \mbox{(e)} & = & 2 G_0(\vr = \vo) \xi^{-2} \int (1 - \rho_0^2)
  \nonumber \\
  & \approx & \frac{1}{2\pi} \ln (1 + 2\pi X) (H[\psi_0] + \pi N)
    + \order(\ta^2/R^2)
  \label{IH}
\eea
where we used (\ref{HBc}), (\ref{DH}) for $m_i = \pm 1$ and $N$ is the total
number of vortices in $\psi_0$.
The higher order loops all give {\em positive} contributions to the vortex
chemical potential.
The most important of them is
\bea
  \half \times \mbox{(f)} & = & \frac{2}{\xi^4} \int_1 \int_2 G_0(1,2)^2
    (1 - \rho_0(1)^2) (1 - \rho_0(2)^2)
  \nonumber \\
  & \lsim & 2 \, \xi^{-2} \left( \int G_0^2 \right) \, \xi^{-2} \int (1 -
\rho_0^2)^2
  \approx \pi N \frac{X}{1 + 2\pi X}
    + \order(\ta^2/R^2) \ ,
  \label{IJ}
\eea
where the $\lsim$ sign expresses that we used a Schwarz inequality to
estimate the integral at the l.h.s., and we used (\ref{HE}), (\ref{CHc}),
(\ref{CJ}) to evaluate the r.h.s. (note that we took the symmetry factor of
the diagram already into account).
Similarly we treat the next diagram,
\bea
  \frac{1}{4} \times \mbox{(g)} & = & \frac{1}{4\xi^4} \int_1 \int_2
G_0(1,2)^2
    (1 - \rho_0(1)^2) (1 - \rho_0(2)^2) \cos 2 (\theta_0(1) - \theta_0(2))
  \nonumber \\
  & \lsim & \frac{\pi N}{8} \frac{X}{1 + 2\pi X}
    + \order(\ta^2/R^2) \ ,
  \label{IK}
\eea
where in addition we neglected the cosine factor on the l.h.s., which is of
course a more serious approximation; however, the contribution of (\ref{IK})
to is smaller than that of (\ref{IJ}) by a factor of 1/8 in any case, so we
neglect this error.
We note that both approximations (\ref{IJ}), (\ref{IK}) would be good in a
limit where the range of $G_0(\vr)^2$ is much smaller than that of $(1 -
\rho_0(\vr)^2)$, but (\ref{HC}), (\ref{CF}) imply that both ranges are
actually of the order of $\xi$.

The dominant class of terms in the whole series is given by loop diagrams of
the form shown in fig.\ \ref{diag3} with $n+1$ nodes, $n$ = 1, 2, ..., and a
symmetry factor of $1/(n+1)$ (the first two are (f) and (h) in fig.\
\ref{diag2}).
An upper limit to the contribution of any of these diagrams may be found by
again using a Schwarz inequality as above, and directly applying
(\ref{HDa}):
\bea
  \lefteqn{ \frac{1}{n+1} \times (\mbox{Fig.\ \ref{diag3}}) =
  } \nonumber \\
  & &\frac{1}{n+1} \left( \frac{2}{\xi^2} \right)^{n+1}
    \int_1 \cdots \int_{n+1} G_0(1,2) \cdots G_0(n+1,1)
    (1 - \rho_0(1)^2) \cdots (1 - \rho_0(n+1)^2)
  \nonumber \\
  & &\quad \lsim \ \frac{1}{2\pi n(n+1)} \, \xi^{-2} \int (1 -
\rho_0^2)^{n+1}
\label{IL}
\eea
The integrals on the r.h.s.\ monotonically tend to zero with increasing $n$
and their coefficients alone form a rapidly converging series with sum
$1/2\pi$ (for comparison, the first term corresponding to diagram (f) in
fig.\ \ref{diag2} already contributes $1/4\pi$).
Since we already overestimated the contributions of (f), (g) we therefore
neglect all higher ``loop" diagrams and hope for the best.
We will see in the next subsection that any further positive contribution to
the vortex chemical potential {\em amplifies} the observed effect and
thereby strengthens our argument anyway.
Putting everything together, we now obtain
\bea
  \tS[\psi_0]
  & = & - \left\{ \frac{K}{2} - \frac{1}{2\pi} \ln (1 + 2\pi X) \right\}
    H[\psi_0] + \frac{N}{2} \ln (1 + 2\pi X)
  \nonumber \\
  & & \quad + 2\pi N \left( \half + \frac{1}{16} \right) \frac{X}{1 + 2\pi
X}
    + \order(\ta^2/R^2) \ ,
  \label{IM}
\eea
which together with (\ref{EM}) gives the effective action of the interacting
vortex system.

We finally note that of course all loop diagrams in figs.\ \ref{diag2},
\ref{diag3} can be calculated numerically to an arbitrary precision, but in
view of the many approximations and qualitative arguments involved in our
derivation of $\tS$ we considered the above rough estimation to be more
adequate.

\subsection{Vortex phase space division and effective Coulomb gas
parameters}

The result (\ref{IM}) of the last subsection already correctly defines the
partition sum (\ref{AF}) of the vortex system.
However, the unit of length in (\ref{AF}) is $\ta$ (which corresponds to
$\Rc$ in eq.\ (\ref{AH})), so that in order to make the link to treatments
of the neutral Coulomb gas in the literature we have to write the effective
action in the form
\be
  \tS[\psi_0]
  = \frac{1}{\TCG} \sum_{i<j} m_i m_j \ln \frac{|\vR_i - \vR_j|}{\ta}
    + N \ln \zCG \ ,
\label{JA}
\ee
whereas in (\ref{EM}) the original lattice spacing $a$ appears in the
logarithm.
(\ref{JA}) then defines the effective dimensionless Coulomb gas temperature
$\TCG$ and fugacity $\zCG$.
Using the neutrality condition $\sum_i m_i = 0$ and the fact that $m_i = \pm
1$ we can rewrite the sum in (\ref{EM}) as
\be
  \sum_{i<j} m_i m_j \ln \frac{|\vR_i - \vR_j|}{a}
  = \sum_{i<j} m_i m_j \ln \frac{|\vR_i - \vR_j|}{\ta}
  - \frac{N}{2} \ln \frac{\ta}{a} \ .
\label{JB}
\ee
The second term will contribute to the fugacity $\zCG$, so we have to
express $\ta$ as a function of the model parameters $a,\xi$.
Recall that $\ta$ was an infrared cutoff on phase fluctuations defined so as
to have the same effect as a mass $2/\xi^2$.
To determine $\ta$ from this condition, note that the most important term in
the perturbation series is (f) in fig.\ \ref{diag2} (and eq.\ (\ref{IA}))
since it is the only one which contributes to the logarithmic vortex
potential.
Its size is mainly determined by the factor
\be
  \frac{K}{2} \langle |\psi_1|^2 \rangle
  = \frac{1}{4\pi} \ln \left( 1 + 2\pi \frac{\xi^2}{a^2} \right)
\label{JC}
\ee
calculated in (\ref{HBc}).
Now if one had chosen the ``hard" momentum cutoff $\pi/\ta$ instead of the
mass, the local fluctuations would have been given by
\be
  \frac{K}{2} \langle |\psi_1|^2 \rangle
  = \int_{{\textstyle \frac{\pi}{\ta}} < k_{x,y} < {\textstyle
\frac{\pi}{a}}}
    \frac{d^2 \vk}{(2\pi)^2} \frac{1}{\vk^2}
  \approx \frac{1}{4\pi} \int _{4\pi/\ta^2}^{4\pi/a^2}
    \frac{d(k^2)}{k^2}
  = \frac{1}{4\pi} \ln \frac{\ta^2}{a^2} \ .
\label{JD}
\ee
Identifying this expression with (\ref{JC}) leads to
\be
  \ln \frac{\ta}{a} = \half \ln \left(1 + 2\pi \frac{\xi^2}{a^2} \right)
\label{JE}
\ee
or, equivalently,
\be
  \ta^2 = a^2 + 2\pi \xi^2 \ .
\label{JF}
\ee
We insert (\ref{JE}) successively in (\ref{JB}), (\ref{EM}), (\ref{IM}) and
obtain the required form (\ref{JA}) of the effective action, the Coulomb gas
parameters now being given by the following expressions:
\bea
  \frac{1}{\TCG} & = & 2\pi K - 2 \ln (1 + 2\pi X) \ ,
  \label{JG} \\
  \ln \zCG & = & \frac{1}{2 \TCG} \left\{
    2 \muXY + F(X) - \half \ln (1 + 2\pi X)
    \right\}
  \nonumber \\
  & & \quad + \half \ln  (1 + 2\pi X) + \frac{9\pi}{8} \frac{X}{1 + 2\pi X}
\ ,
\label{JH}
\eea
where $2\muXY = -1.617$ (see eq.\ (\ref{EB})), $X = \xi^2/a^2$, and $F(X)$
is defined by eq.\ (\ref{EE}).

Equations (\ref{JG}), (\ref{JH}) are the final results of this paper.
The expression in curly brackets may be interpreted as a renormalized vortex
chemical potential, whereas the second line of (\ref{JH}) contributes to the
renormalized phase space division.

In fig.\ \ref{CGf} the $\zCG$ vs.\ $\TCG$ curves are drawn in Minnhagen's
Coulomb gas phase diagram for different values of the GL correlation length
$\xi$.
With increasing $\xi/a$ the curves cross the phase boundary at increasingly
higher values of $\zCG$, finally reaching the first order regime for values
of $\xi \gsim a$.
Since we argued in subsection 2.1 that in superconducting films {\em always}
$\xi \gsim a$, we conclude that the latter are good candidates for a first
order vortex unbinding transition.
Because of the approximations and relatively rough estimations involved in
our calculations we do not insist in the {\em quantitative} information
conveyed by fig.\ \ref{CGf}.
However, we believe that the {\em trend} is clear enough to be reliable.

Another interesting (although not unexpected) result might be that for
increasing $\xi/a$ the transition occurs at increasingly higher values of
$K$ (which according to (\ref{ZC}) corresponds to smaller values of the
``physical" temperature).
In other words, the distance between $\Tv$ and $\Tco$ increases with
increasing $\xi/a$.

\section{Summary and conclusions}

In this paper, we investigated the nature of the transition in the two-
dimensional Ginzburg-Landau model of a neutral superfluid.
In doing this, we assumed --- as is usually done --- that the transition is
caused by the collective unbinding of vortex pairs.
However, we paid particular attention to short-wavelength amplitude and
phase fluctuations of the order parameter on length scales $\lsim \xi$ (the
Ginzburg-Landau correlation length), which we showed to be strongly coupled
to the vortices.
This is in contrast with previous work in the literature where amplitude
fluctuations usually are neglected altogether and phase fluctuations are
assumed to be only weakly coupled to the vortices.

Eliminating perturbatively these short-wavelength fluctuations, we derived
an effective free energy for the vortex degrees of freedom.
We argued that this effective vortex gas still is a 2D Coulomb gas (i.e.,
the vortex-vortex interaction varies logarithmically at large distances),
however both the effective temperature and the fugacity of the Coulomb gas
are strongly  renormalized if $\xi$ is larger than a microscopic cutoff
length $a$ for fluctuations.
We argued that these considerations should be relevant for those
superconducting films for which a BCS description is qualitatively correct,
since BCS theory predicts that {\em always} $\xi \gsim a$.

By this elimination process of small-scale fluctuations we furthermore
obtained a microscopic interpretation of the effective vortex phase space
division $\Delta$ as an interesting secondary result.
In particular, we could clarify the relation between $\Delta$ and the
Ginzburg-Landau correlation length $\xi$.

Our subsequent conclusions concerning a possible first order transition then
relied completely on the correctness of Minnhagen's self-consistent theory
of a dense 2D Coulomb gas, which predicts a first order vortex unbinding
transition with non-universal (not KT like) properties for sufficiently
large values of the vortex fugacity ($\zCG \gsim 0.05$).
Unfortunately, the rather subtle differences between a KT like and a first
order vortex unbinding transition seem to be outside the scope of present
experimental investigations of real superconducting films.
However, we think that an interesting and probably feasible (though likewise
very hard) problem would be to numerically investigate the critical
properties of a lattice Ginzburg-Landau model including fluctuating order
parameter amplitudes.

\subsection*{Acknowledgements}

We thank D. Ariosa, D. Baeriswil, U. Eckern, F. Mila and P. Minnhagen for
clarifying discussions.
One of us (D. B.) gratefully acknowledges financial support by the Swiss
National Science Foundation.

\newpage

\section*{Tables}

\begin{table}[h]
  \begin{tabular}{|c|c|c|c|} \hline
  model & $\Rc$     & $\Delta$     & $\muCG$ \\ \hline\hline
  XY    & $a$       & $a^2$        & -0.809  \\ \hline
  GLCG  & $2.24\xi$ & $16.4\xi^2$  & -0.390  \\ \hline
  \end{tabular}\\[.5cm]
Table I: Effective CG parameters for the XY model (values from ref.\
\protect\cite{JKK77}) and for the ``bare" GLCG (values from refs.\
\protect\cite{MN85,WM88}).
\end{table}

\begin{table}[h]
  \begin{tabular}{|c|c|c|c|} \hline
  $m$ & $E_1$  & $E_2$ & $E_3$ \\ \hline\hline
  1   & -0.806 & 0.279 & 1     \\ \hline
  2   &        & 0.416 & 4     \\ \hline
  \end{tabular}\\[.5cm]
Table II: Some numerical values for the $E$ parameters of the 1-vortex
problem (eqs.\ (\protect\ref{CHa}), (\protect\ref{CHb}),
(\protect\ref{CHc})), taken from refs.\ \protect\cite{Hu72,MN85}.
See explanation in the text.
\end{table}

\newpage

\section*{Figure Captions}

\begin{figure}[h]
\caption{\label{CG1}
Plot of Minnhagen's generic CG phase diagram \protect\cite{Min85,TK88,MW89}.
Also shown are the $\zCG(\TCG)$ relations (eq.\ (\protect\ref{AI})) for the
XY and GLCG models, using the parameters of table I.
In both cases, a KT like vortex unbinding transition is predicted.
}
\end{figure}

\begin{figure}[h]
\caption{\label{CG2}
Same as fig.\ \protect\ref{CG1}, but with several different values of the
cutoff $\Rc$ of the vortex-vortex interaction (see eq.\ (\protect\ref{AH})).
The qualitative properties of the vortex unbinding transition of the XY and
GLCG models are not changed.
}
\end{figure}

\begin{figure}[h]
\caption{\label{diag1}
Diagrams corresponding to the interaction term $H_I$ of the hamiltonian,
eq.\ (\protect\ref{IA}).
}
\end{figure}

\begin{figure}[h]
\caption{\label{diag2}
Diagrams corresponding to the gaussian approximation of the effective action
$\tS[\psi_0]$.
}
\end{figure}

\begin{figure}[h]
\caption{\label{diag3}
Class of diagrams represented in eq.\ (\protect\ref{IL}).
}
\end{figure}

\begin{figure}[h]
\caption{\label{CGf}
Again Minnhagen's CG phase diagram as in figs.\ \protect\ref{CG1},
\protect\ref{CG2}, but now with our final $\zCG(\TCG)$ relation for the GLCG
(eqs.\ (\protect\ref{JG}), (\protect\ref{JH})), plotted for different values
of $\xi/a$.
The limit $\xi/a = 0$ corresponds to the XY model as shown in fig.\
\protect\ref{CG1}.
With increasing $\xi/a$ the $\zCG(\TCG)$ line is shifted towards the upper
part of the phase diagram, finally reaching the first order part of the
vortex unbinding transition.
}
\end{figure}

\clearpage


\newpage

\begin{thebibliography}{99}

\bibitem{ODLRO} O. Penrose, L. Onsager: Phys.\ Rev.\ {\bf 104}, 576 (1956);
C. N. Yang: Rev.\ Mod.\ Phys.\ {\bf 34}, 694 (1962)
\bibitem{2DLRO} N. D. Mermin, H. Wagner: Phys.\ Rev.\ Lett.\ {\bf 17}, 1133
(1966); P. C. Hohenberg: Phys.\ Rev.\ {\bf 158}, 383 (1967)
\bibitem{Las68} G. Lasher: Phys.\ Rev.\ {\bf 172}, 224 (1968)
\bibitem{FBJ73} M. E. Fisher, M. N. Barber, D. Jasnow: Phys.\ Rev.\ A {\bf
8}, 1111 (1973)
\bibitem{BR80} D. J. Bishop, J. D. Reppy: Phys.\ Rev.\ B {\bf 22}, 5171
(1980)
\bibitem{BMO79} M. R. Beasley, J. E. Mooij, T. P. Orlando: Phys.\ Rev.\
Lett.\ {\bf 42}, 1165 (1979)
\bibitem{HN79} B. I. Halperin, D. R. Nelson: J. Low Temp.\ Phys.\ {\bf 36},
599 (1979)
\bibitem{HF80} A. F. Hebard, A. T. Fiory: Phys.\ Rev.\ Lett.\ {\bf 44}, 291
(1980)
\bibitem{SCfilms} A. F. Hebard, A. T. Fiory: Phys.\ Rev.\ Lett.\ {\bf 50},
1603 (1983); A. T. Fiory, A. F. Hebard, W. I. Glaberson: Phys.\ Rev.\ B {\bf
28}, 5075 (1983)
\bibitem{LFM90} C. Leemann, P. Fl\"uckiger, V. Marsico, J. L. Gavilano, P.
K. Srivastava, P. Lerch, P. Martinoli: Phys.\ Rev.\ Lett.\ {\bf 64}, 3082
(1990)
\bibitem{VLV91} S. Vadlamannatti, Q. Li, T. Venkatesan, W. L. McLean, P.
Lindenfeld: Phys.\ Rev.\ B {\bf 44}, 7094 (1991)
\bibitem{Ber71} V. L. Berezinski\u{\i}: Zh.\ Eksp.\ Teor.\ Fiz.\ {\bf 61},
1144 (1971) [Sov.\ Phys.\ JETP {\bf 34}, 610 (1972)]
\bibitem{KT72} J. M. Kosterlitz, D. J. Thouless: J. Phys.\ C {\bf 5}, L124
(1972)
\bibitem{KT73} J. M. Kosterlitz, D. J. Thouless: J. Phys.\ C {\bf 6}, 1181
(1973)
\bibitem{Nel83} D. R. Nelson, in: {\em  Phase Transitions and Critical
Phenomena}, Vol.\ 7, edited by C. Domb and J. L. Lebowitz (Academic Press,
London, 1983), p.\ 1
\bibitem{BCSGL} N. R. Werthamer, in: {\em Superconductivity}, ed. R. D.
Parks (Marcel Dekker, New York, 1969), p.\ 321; M. Cyrot: Rep.\ Prog.\
Phys.\ {\bf 36}, 103 (1973)
\bibitem{Hal79} B. I. Halperin, in: {\em Physics of low-dimensional
Systems}, Proceedings of the Kyoto Summer Institute 1979, edited by Y.
Nagaoka and S. Hikami (Publication Office, Progr.\ Theor.\ Phys., Kyoto,
1979), p.\ 53
\bibitem{vortexloops} R. P. Feynman: Prog.\ Low Temp.\ Phys.\ {\bf 1}, 52
(1955); G. A. Williams: Phys.\ Rev.\ Lett.\ {\bf 59}, 1926 (1987); Phys.\
Rev.\ Lett.\ {\bf 68}, 2054 (1992); S. R. Shenoy: Phys.\ Rev.\ B {\bf 40},
5056 (1989)
\bibitem{Min87} P. Minnhagen: Rev.\ Mod.\ Phys.\ {\bf 59}, 1001 (1987)
\bibitem{KTRG} J. M. Kosterlitz: J. Phys.\ C {\bf 7}, 1046 (1974); P. B.
Wiegmann: J. Phys.\ C {\bf 11}, 1583 (1978); D. J. Amit, Y. Y. Goldstein, G.
Grinstein: J. Phys.\ A {\bf 13}, 585 (1980)
\bibitem{JKK77} J. V. Jos\'e, L. P. Kadanoff, S. Kirkpatrick, D. R. Nelson:
Phys.\ Rev.\ B {\bf 16}, 1217 (1977);
\bibitem{NK77} D. R. Nelson, J. M. Kosterlitz: Phys.\ Rev.\ Lett.\ {\bf 39},
1201 (1977)
\bibitem{MW81} P. Minnhagen, G. G. Warren: Phys.\ Rev.\ B {\bf 24}, 2526
(1981)
\bibitem{JN91} W. Janke, K. Nather: Phys.\ Lett.\ A {\bf 157}, 11 (1991)
\bibitem{JMN93} A.\ Jonsson, P.\ Minnhagen, M.\ Nyl\'en: Phys.\ Rev.\ Lett.\
{\bf 70}, 1327 (1993) 
\bibitem{CGnumer} J. M. Caillol, D. Levesque: Phys.\ Rev.\ B {\bf 33}, 499
(1986); J.-R. Lee, S. Teitel: Phys.\ Rev.\ Lett.\ {\bf 64}, 1483 (1990);
Phys.\ Rev.\ Lett.\ {\bf 66}, 2100 (1991)
\bibitem{JJAs} C. Leemann, P. Lerch, G. A. Racine, P. Martinoli: Phys.\
Rev.\ Lett.\ {\bf 56}, 1291 (1986); P. Martinoli, P. Lerch, C. Leemann, H.
Beck: Proc.\ of the {\em 18th Conference on Low Temperature Physics}, Kyoto
1987, Japanese Journal of Applied Physics {\bf 26}, 1999 (1987)
\bibitem{JFG89} B. Jeanneret, P. Fl\"uckiger, J. L. Gavilano, C. Leemann, P.
Martinoli: Phys.\ Rev.\ B {\bf 40}, 11374 (1989)
\bibitem{Pel78} R. Pelcovits: Ph.\ D.\ Thesis, Harvard University, 1978
(unpublished)
\bibitem{MN85} P. Minnhagen, M. Nyl\'en: Phys.\ Rev.\ B {\bf 31}, 5768
(1985)
\bibitem{Min85} P. Minnhagen: Phys.\ Rev.\ B {\bf 32}, 3088 (1985)
\bibitem{TK88} J. M. Thyssen, H. J. F. Knops: Phys.\ Rev.\ B {\bf 38}, 9080
(1988)
\bibitem{MW89} P. Minnhagen, M. Wallin: Phys.\ Rev.\ B {\bf 40}, 5109 (1989)
\bibitem{You78} A. P. Young: J. Phys.\ C {\bf 11}, L453 (1978)
\bibitem{WM88} H. Weber, P. Minnhagen: Phys.\ Rev.\ B {\bf 38}, 8730 (1988)
\bibitem{Mil93} F. Mila: Phys.\ Rev.\ B {\bf 47}, 442 (1993)
\bibitem{Gor59} L. P. Gor'kov: Zh.\ Eksp.\ Teor.\ Fiz.\ {\bf 36}, 1918
(1959) [Sov.\ Phys.\ JETP {\bf 9}, 1364 (1959)]
\bibitem{Pea64} J. Pearl: Appl.\ Phys.\ Lett.\ {\bf 5}, 65 (1964)
\bibitem{Hu72} C. R. Hu: Phys.\ Rev.\ B {\bf 6}, 1756 (1972)

\end{thebibliography}
\end{document}